\documentclass[prd,twocolumn,amsmath,amssymb,floatfix,superscriptaddress,nofootinbib]{revtex4-1}
\usepackage{bm}
\usepackage{amsmath}
\usepackage{epsfig}
\usepackage{color}
\usepackage{natbib}
\usepackage{textcase}
\usepackage{graphicx}
\usepackage{ifthen}
\usepackage{xstring}
\usepackage{graphicx}
\usepackage[utf8]{inputenc}

\newcommand{\refeq}[1]{Eq.~(\ref{eq:#1})}

\newcommand{\reffig}[1]{Fig.~\ref{fig:#1}}

\newcommand{\xef}{x_e^{\rm fid}}

\newcommand{\zmax}{z_{\rm max}}

\newcommand{\xemin}{x_e^{\rm min}}

\newcommand{\beq}{\begin{equation}}
\newcommand{\eeq}{\end{equation}}

\newcommand{\bea}{\begin{eqnarray}}
\newcommand{\eea}{\end{eqnarray}}

\definecolor{darkgreen}{cmyk}{0.85,0.2,1.00,0.2} 
\definecolor{purple}{cmyk}{0.5,1.0,0,0}

\definecolor{ultramarine}{rgb}{0.07, 0.04, 0.56}
\definecolor{cadmiumgreen}{rgb}{0.0, 0.42, 0.24}
\definecolor{indigo(dye)}{rgb}{0.0, 0.25, 0.42}
\usepackage[linktocpage=true]{hyperref}
\hypersetup{
colorlinks=true,
citecolor=ultramarine,
linkcolor=cadmiumgreen,
urlcolor=indigo(dye),
pdfauthor={},
pdftitle={},
pdfsubject={}
}

\begin{document}
	
\title{Does Planck 2015 polarization favor high redshift reionization?}

\author{Chen Heinrich}\email{chenhe@caltech.edu}
\affiliation{Jet Propulsion Laboratory, California Institute of Technology, Pasadena California 91109}

\author{Wayne Hu}
\affiliation{Kavli Institute for Cosmological Physics, Enrico Fermi Institute, University of Chicago, Chicago Illinois 60637}
\affiliation{Department of Astronomy \& Astrophysics,
 University of Chicago, Illinois 60637}

\begin{abstract}
We study the relationship between signatures of high redshift ionization in large-angle CMB polarization power spectra and features in the Planck 2015 data.   Using a principal component (PC) ionization basis
that is complete to the cosmic variance limit out to $\zmax=30,40,50$, we find a robust
$>95\%$ CL preference for ionization at $z>15$ with no preference for $z>40$.  
This robustness originates from the $\ell \sim 10$ region of the data which show high power 
relative to $\ell \le 8$ and result in a poor fit to a steplike model of reionization.   Instead
by allowing for high redshift reionization, the PCs provide a better fit by $2\Delta \mathrm{ln}\mathcal{L} = 5-6$.   Due to a degeneracy in the ionization redshift response, this improved fit is due to a single aspect of the model: the ability to accommodate  $z>10$ component to the ionization as we
illustrate with a two-step reionization model.
 For this and other models that accommodate such a component, its presence is allowed and even favored; for models that do not, their poor fit reflects statistical or systematic
fluctuations.   These possibilities produce very different and testable predictions at $\ell \sim 15-20$, as well as small but detectable differences at $\ell>30$ that can further restrict the
high redshift limit of reionization.
\end{abstract}
\maketitle

\section{Introduction}

The detailed process of reionization remains one of the least well-understood aspect of the standard model of cosmology (see e.g.~\cite{2016ASSL..423.....M, 2018ASSL} and references therein) and yet its modeling has implications for many other cosmological inferences from the cosmic microwave background (CMB) \cite{Ade:2015xua}, such as the initial power spectrum \cite{Mortonson:2009xk,Mortonson:2009qv, 2012arXiv1205.0463T}, the
growth of structure~\cite{Hu:2003pt}, and the sum of the neutrino masses~\cite{Smith:2006nk,Allison:2015qca, Watts:2018etg}.

The standard approach to modeling the impact of reionization on the CMB is through an averaged global ionization history corresponding to a sudden steplike transition to fully ionized hydrogen and singly ionized helium. This  model assumes a priori that there is negligible ionization at high redshift before the transition. Yet the shape of the reionization bump in the large-angle CMB $E$-mode polarization carries coarse grained information about the redshift evolution of the ionization history, and can in principle reveal information for high redshift ionization that requires more than 
a steplike transition.

Indeed, Refs.~\cite{Hu:2003gh,Mortonson:2007hq} developed a principal component  (PC)  based technique to characterize the redshift information 
embedded in the $EE$ power spectrum out to a given maximum redshift $\zmax$ to the cosmic variance limit.
Recently, an implementation of this technique on the Planck 2015 data out to $\zmax=30$ 
revealed that ionization at $z>15$ was not only allowed but preferred at $>95\%$ CL  \cite{Heinrich:2016ojb}.  Using the effective likelihood of the 5 PC parameters developed and released in 
Ref.~\cite{Heinrich:2016ojb}, one can assess the implications for {\it any} model of reionization out to $\zmax=30$ given the completeness
property of the PC description (see e.g.~\cite{Miranda:2016trf}).

Two sets of related questions arise from this study.\footnote{A third question, whether the polarization features that drive reionization constraints are related to or affected by the known large angle temperature anomalies, is addressed in a separate work  \cite{Obied:2018qdr}.}     The first is what aspect of the Planck  2015 data drives this preference for high redshift ionization
 and how can improved measurements in the final Planck release or future measurements 
test these inferences.  The second is the impact of choosing a PC parameterization to $\zmax=30$.   Does adding these specific extra parameters simply fit expected random statistical fluctuations in the measurements multipole to multipole?  Does the choice of  PC parameters
with its enhanced parameter volume out to $\zmax$ introduce a prior preference for high redshift ionization that increases as $\zmax$ increases?
More generally how does one remove any biases from the PC prior in the context of a physical model for reionization
(cf.~\cite{Villanueva-Domingo:2017ahx})?

In this paper, we address these questions.   We begin in Sec.~\ref{sec:method} with a study of 
the detailed response of the power spectrum observables to ionization at a given redshift to see how and why they fit
the Planck polarization data.   We then conduct in Sec.~\ref{sec:results} a PC analysis of the Planck 2015 data increasing $\zmax$ to 40 and 50
to study the robustness of the implications for high redshift ionization.  We discuss the results in Sec.~\ref{sec:discussion}.
In Appendix \ref{sec:appendix} we provide worked examples of the impact of the PC prior on our results and
the encapsulation of the improved fit into a single parameter that represents the high redshift optical depth.

\section{Reionization observables}\label{sec:method}

\begin{figure}
          \includegraphics[width=1\columnwidth]{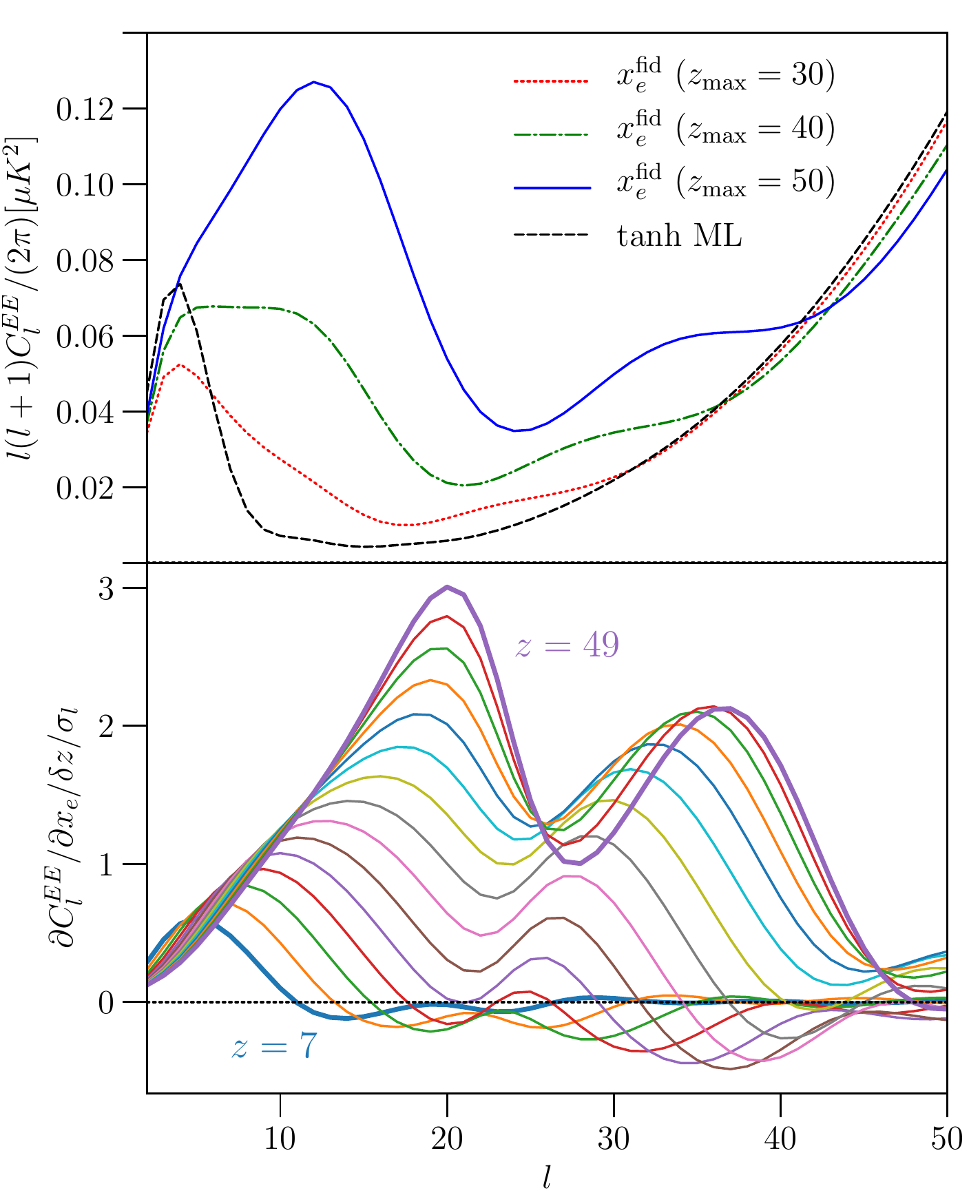}
          \caption 
          {Top: $C_\ell^{EE}$ of fiducial models around which the PCs for $\zmax=30,40,50$ vary vs.~the tanh chain maximum likelihood model. Bottom: Responses of the $C_\ell^{EE}$ spectrum to the ionization fraction at different redshifts (7 to 49 with spacing $\Delta z = 3$).  Responses are calculated per unit redshift from  $\delta z = 0.25$  perturbations around the fiducial $\zmax=50$ model  
          and scaled to its cosmic variance per multipole.   Note the similarity of the $\ell \lesssim 10$ responses for $z\gtrsim 10$ and differences at $\ell \gtrsim 15$  and $\ell \gtrsim 30$.}
            \label{fig:dClEE_dxe}
\end{figure}

\begin{figure}
          \includegraphics[width=1\columnwidth]{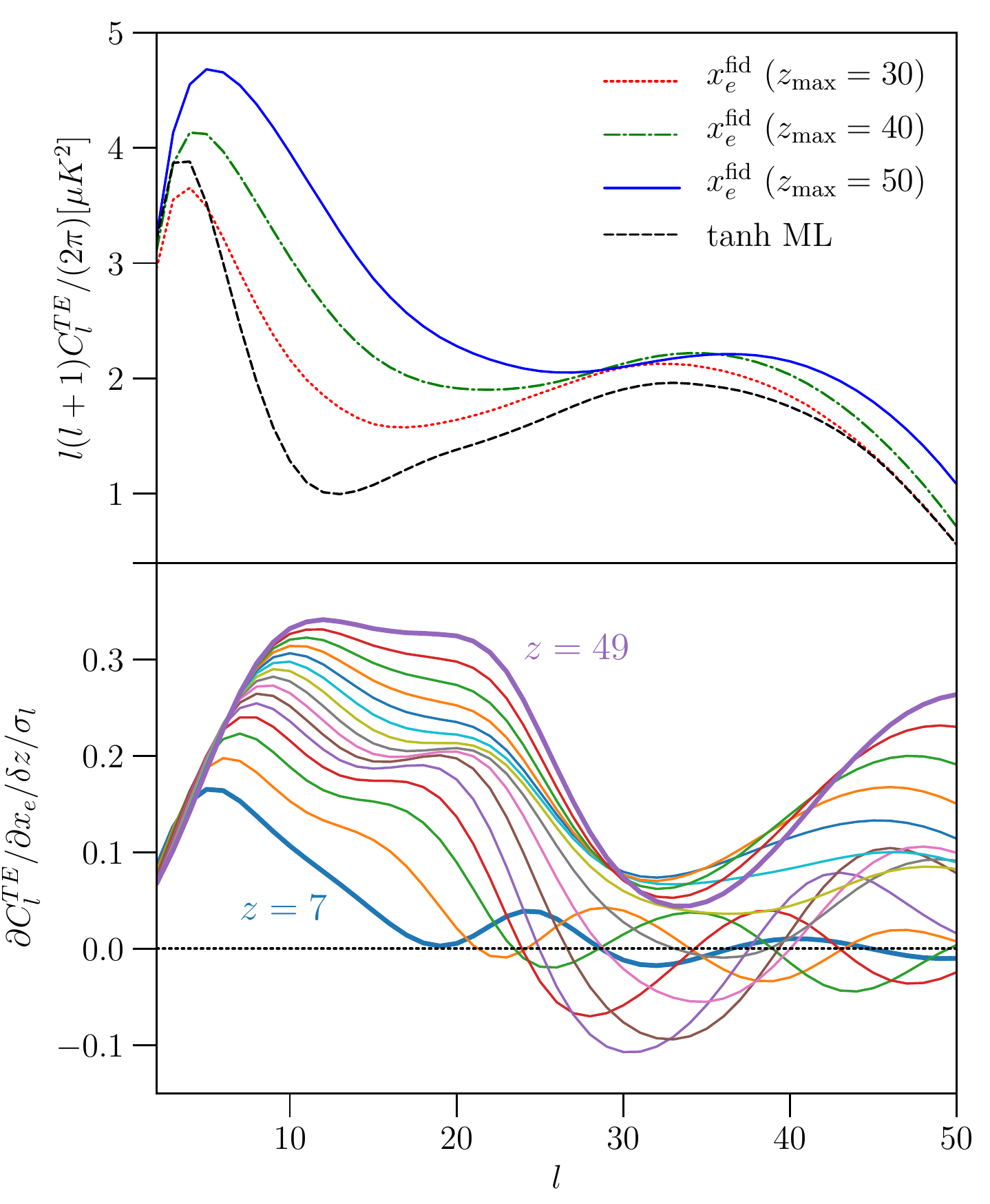}
          \caption
          {Same as \reffig{dClEE_dxe} but for $C_\ell^{TE}$. Note again the similarity of the $\ell \lesssim 10$ responses for $z\gtrsim 10$. }
            \label{fig:dClTE_dxe} 
\end{figure}

The observable impact  on the large-angle $E$-mode polarization spectrum of any ionization history out to a given $z_{\rm max}$ in the reionization range can be completely characterized  to cosmic variance precision by just a few principal component parameters.  We mainly follow Ref.~\cite{Heinrich:2016ojb} for the PC construction but highlight the role of  $z_{\rm max}$ and its relationship to features in the observable power spectra.

We begin by allowing arbitrary perturbations of the  ionization fraction relative to the fully ionized hydrogen
$x_e(z)$ around a fiducial model.   
Following Ref.~\cite{Heinrich:2016ojb}, the fiducial model is $\xef(z) = 0.15$ in the range $z_{\rm min}<z<z_{\rm max}$ and vanishes for $z>z_{\rm max}$ (see example of $\zmax = 30$ in Ref.~\cite{Heinrich:2016ojb}, Fig.~1). We take $z_{\rm min}=6$ and assume fully ionized hydrogen and singly ionized helium, $\xef(z) = 1+ f_{\rm He}$ for $z_{\rm He} \lesssim z<z_{\rm min}$, consistent with Ly$\alpha$ forest constraints (e.g.\ \cite{Becker:2015lua}). Helium becomes fully ionized at $z_{\rm He}$, which is modeled here  as a tanh function in redshift centered at  $z_{\rm He} = 3.5$ \cite{Becker:2010cu} with width $\Delta z=0.5$.  Therefore  $x_e =  1+ 2f_{\rm He}$ for $z \lesssim z_{\rm He}$. Here,
\beq
f_{\rm He} = \frac{n_{\rm He}}{n_{\rm H}} = \frac{m_{\rm H}}{m_{\rm He}}\frac{Y_p}{1-Y_p}
\eeq
is the ratio of the helium to hydrogen number density, and $Y_p$ is the helium mass fraction, set to be consistent with big bang nucleosynthesis for the chosen baryon density. 
The $EE$ and $TE$ power spectra for the fiducial model with $z_{\rm max}=30,40,50$ are shown in Figs.~\ref{fig:dClEE_dxe} and~\ref{fig:dClTE_dxe} (upper panels).

These fiducial models differ substantially from each other and from the
standard tanh or steplike reionization model
 \begin{equation}
x_e^{\rm true}(z) = \frac{1+f_{\rm He}}{2}\left\{  1+ \tanh\left[ \frac{y(z_{\rm re})-y(z)}{\Delta y} \right] \right\},
 \label{eqn:tanh}
 \end{equation}
 with $y(z)=(1+z)^{3/2}$, $\Delta y=(3/2)(1+z)^{1/2}\Delta z$, and $\Delta z = 0.5$.  
This tanh model is also shown in Figs.~\ref{fig:dClEE_dxe} and~\ref{fig:dClTE_dxe} (black dashed) for the best fit $z_{\rm re}= 9.85$ ($\tau = 0.0765$) from Ref.~\cite{Heinrich:2016ojb}.
 Note in particular the very different shape of the
 $EE$ power spectrum which has a sharper peak at lower multipoles than all the fiducial models.

Next we consider variations around these fiducial models.   In practice, we discretize $x_e(z)$ with a redshift spacing $\delta z = 0.25$ so that each $x_e(z_i)$ is a parameter.   The full 
ionization history is constructed as the linear interpolation of these discrete perturbations.
In Figs.~\ref{fig:dClEE_dxe} and \ref{fig:dClTE_dxe} (lower panel), we display the observable responses to perturbations in
$x_e(z_i)$ around the $\zmax=50$ fiducial model through the derivatives $\partial C_\ell^{EE}/\partial x_e(z_i)/\delta z$ and $\partial C_\ell^{TE}/\partial x_e(z_i)/\delta z$ respectively. To highlight the ultimate observability of these variations, we scale the derivatives to the cosmic variance per $\ell$ mode 
  \begin{equation}
 \sigma_{\ell} =
 \begin{cases}
\sqrt{\frac{1}{2\ell+1}}  \sqrt{ C_\ell^{TT} C_\ell^{EE} + (C_\ell^{TE})^2}, & TE;\\
\sqrt{\frac{2}{2\ell+1}}  C_\ell^{EE}, & EE. \\
\end{cases}
\end{equation}
This scaling highlights the fact that most of the information on the ionization history will ultimately 
come from the $EE$ spectrum even though for the low signal to noise of the Planck polarization measurement $TE$ contributes as well. 

Relatedly, by scaling to the cosmic variance rather than the Planck noise variance, we visually overweight the low signal-to-noise regions.   For this reason when displaying the data and power spectrum differences,
  we also choose to calculate $\sigma_\ell$ with the fiducial model for $z_{\rm max} = 50$ rather than
  the tanh model which has an even smaller signal.  We retain this convention for the
  $\zmax=30,40$ cases so that power spectrum differences and data are displayed with the same convention.
  To compute power spectra, we use a modified version of CAMB\footnote{CAMB: \url{http://camb.info}}~\cite{Lewis:1999bs, Howlett:2012mh} and to calculate derivatives 
  we use a double sided
finite difference of fixed optical depth $\delta \tau\approx \pm 0.0006$ and 0.001 for $\zmax = 40$ and 50 respectively  to assure convergence and numerical
stability. These derivatives are calculated at fixed $A_s e^{-2\tau}$ rather than fixed power spectrum normalization $A_s$ since the former is well constrained from the acoustic peaks. 
For $\zmax=30$, we follow the construction from Ref.~\cite{Heinrich:2016ojb}.
  For the calculation of derivatives we boost the accuracy of CAMB and use a small
smoothing in $z$ for numerical stability and accuracy.   For all other calculations we smooth the ionization history
 with a Gaussian in $\mathrm{ln}(1+z)$ of width $\sigma_{\mathrm{ln}(1+z)} = 0.015$
 which speeds up the analysis with negligible loss in precision for realistically smooth ionization histories.

 First notice that to get substantial changes in $EE$ power at $\ell \gtrsim 10$, we require 
 $z_i \gtrsim 10$.  Note also that the relatively large features above $\ell \sim 15$ are influenced by the scaling to the cosmic variance of the
 fiducial model (see Fig.~\ref{fig:dClEE_dxe} top panel) and are in a regime which is not  well measured by Planck.
 For $TE$, the contributions relative to cosmic variance look slightly smoother in multipole given the smooth temperature spectrum though again enhancements at $\ell \gtrsim 10$ require
 $z_i \gtrsim 10$.

  Next notice that the  various perturbations in $x_e(z_i)$ at  $z_i \gtrsim 15$ are very similar in $EE$ for the $\ell \lesssim 10$ regime that is constrained by the data.   They differ more
strongly in the low signal-to-noise region beyond $\ell \sim 15$. 
Given that the quadrupolar sources of polarization are associated with the horizon scale, the higher the redshift the higher the
maximum multipole of the response but even high redshift perturbations affect $\ell \lesssim 10$.   
We shall see that this implies that the best measured regime does not provide a strong constraint on  the specific redshift
at which there is a preference for a finite high redshift ionization fraction.
More detailed 
differentiation requires detections in the low signal regime.

The degenerate responses in the observables to neighboring $z_i$ are also what makes the PC parameterization
much more efficient than a direct redshift space exploration.  We construct the PCs as
the eigenfunctions of the Fisher matrix of the $x_e(z_i)$ perturbations  for  cosmic variance limited $C_\ell^{EE}$ measurements which, as discussed above, ultimately contain almost all of the low multipole information
on reionization
\begin{equation} 
F_{ij} = \sum_\ell \frac{1}{\sigma_\ell^2}  {\partial C_\ell^{EE} \over \partial x_e(z_i)}{\partial C_\ell^{EE} \over \partial x_e(z_j)} = \sum_{a} S_{a}(z_i) \sigma_{a}^{-2} S_{a}(z_j).
\label{eq:fisher}
\end{equation}
Note that in this construction, as opposed to the figures, $\sigma_\ell$ is the cosmic variance per multipole of the fiducial model in question rather than the fixed case of $\zmax=50$.
We again linearly interpolate  the discrete $S_a(z_i)$ to obtain the continuous functions $S_a(z)$ and characterize the ionization history as 
\begin{equation}
x_e(z)=\xef(z)+\sum_{a}m_{a}S_{a}(z),
\label{eq:mmutoxe}
\end{equation}
where $m_a$ are the PC amplitudes.

\begin{figure}
          \includegraphics[width=\columnwidth]{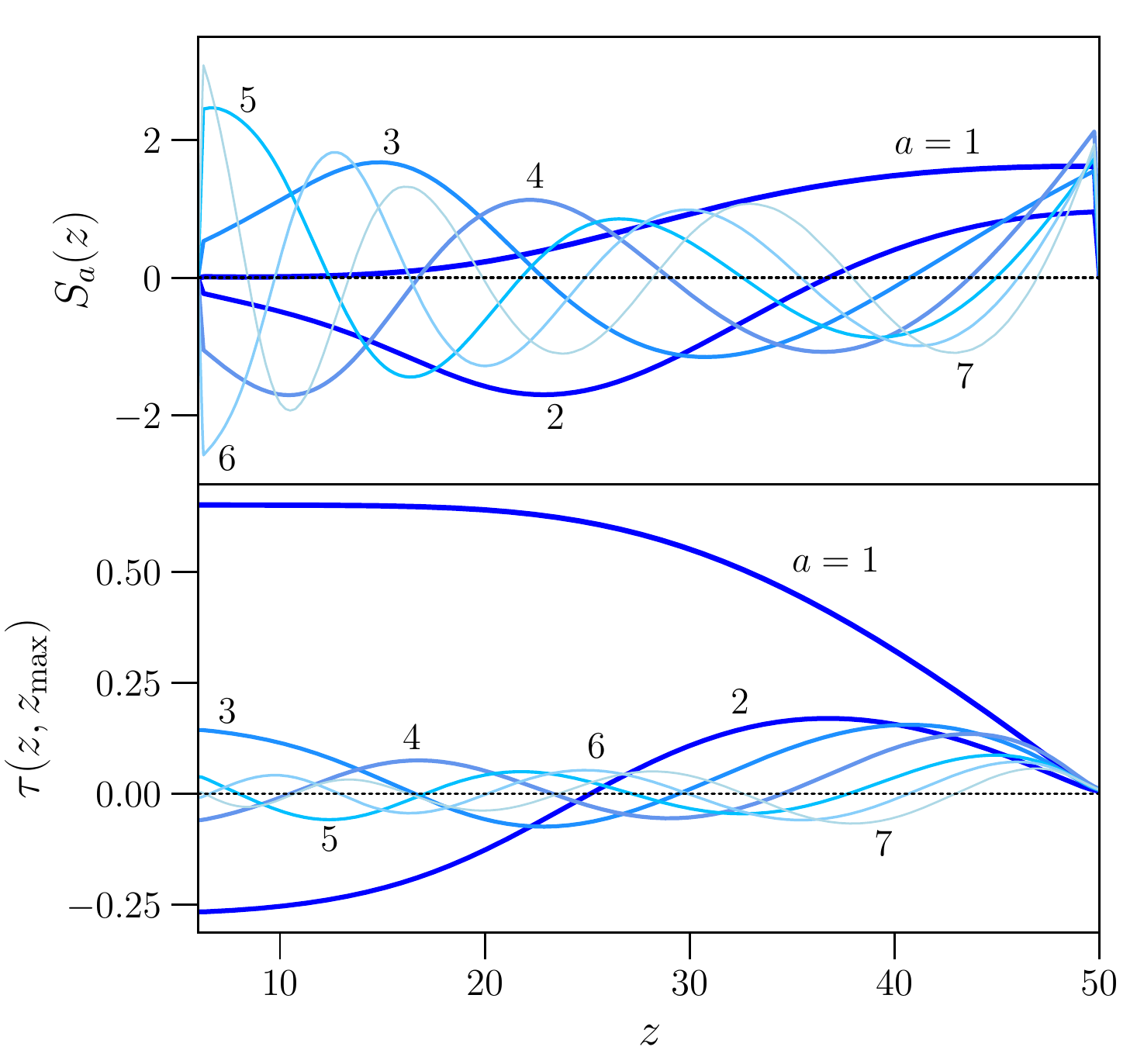}
          \caption{First $n_{\rm{PC}} = 7$ PCs which form a complete  basis for ionization histories up to $\zmax = 50$ with respect to the large angle $C_\ell^{EE}$ observables. Top: Ionization PCs in redshift.  Bottom: Cumulative optical depth for unit amplitude PCs. Higher PCs correspond to finer variations in redshift space which leave unobservable effects. }
          \label{fig:PC_zmax50}
\end{figure}

After obtaining $S_a(z)$ from \refeq{fisher}, we rank order them from the smallest to largest variance $\sigma_a^2$. Then we determine $n_{\rm PC}$, the minimum number  of PCs needed to completely describe the impact on $C_\ell^{EE}$ from any ionization history to cosmic variance limit~\cite{Hu:2003gh}. We find that $n_{\rm PC}$  = 5 suffices for $z_{\rm max}=30$ and 40, whereas $n_{\rm PC} = 7$ is needed for $z_{\rm max}=50$, because of the larger amount of information contained in the larger redshift range that the CMB is able to probe.
Note that despite the large differences in the fiducial models shown in Fig.~\ref{fig:dClEE_dxe}, all three sets of PCs can describe
the tanh model to the cosmic variance limit.

As an illustration, the first seven principal components for $z_{\rm max} = 50$ are shown in the top panel of \reffig{PC_zmax50}. The higher variance PCs correspond to faster oscillations in redshift space.  Consequently their impact  on the cumulative Thomson optical depth 
\begin{equation}
\tau(z,z_{\rm max}) = n_{\rm H}(0) \sigma_T \int_z^{z_{\rm max}} dz \frac{x_e(z) (1+z)^{2} }{H(z)} ,
\label{eq:cumtau}
\end{equation}
is reduced and so $\tau(z,z_{\rm max})$ provides a better visualization of the constraints than
$x_e(z)$.  
Here $n_{\rm H}(0)$ is the hydrogen number density at $z=0$ and $H(z)$ is the expansion rate.
Note that in practice, the integral boundary in $\tau(z,z_{\rm max})$ goes slightly past the nominal $\zmax$ to capture the effect of smoothing $x_e$ beyond $\zmax$.
 In the lower panel of 
\reffig{PC_zmax50},  we show this quantity for the individual PCs with unit amplitude $m_a=1$ and
$z_{\rm max} = 50$.  Since the optical depth as a function of redshift controls the observable
properties of the power spectra, we can see why the first few PCs contain most of the
information for the Planck data.

\section{Reionization PC Constraints}\label{sec:results}

We analyze the Planck data for PC parameterized reionization histories that are complete to
$z_{\rm max}=40,50$ extending the $z_{\rm max}=30$ range 
 in Ref.~\cite{Heinrich:2016ojb}, where the details of the procedure are covered.  
 In brief, we use the Planck public likelihoods plik\_lite\_TTTEEE\footnote{We tested in Ref.~\cite{Heinrich:2016ojb} that results are robust to using the plik\_lite or the full Planck likelihood, so we chose plik\_lite for speed.} for high-$\ell$'s and lowTEB for low-$\ell$'s, which includes the 2015 LFI but not HFI polarization~\cite{Aghanim:2015xee}. To sample the posterior distribution of the PC amplitudes $m_a$, we use the Markov Chain Monte Carlo (MCMC) technique with a modified version of COSMOMC\footnote{COSMOMC: \url{http://cosmologist.info/cosmomc}}~\cite{Lewis:2013hha, Lewis:2002ah}.  We also marginalize the standard  $\Lambda$CDM cosmological parameters: baryon density $\Omega_{b}h^2$,  cold dark matter density $\Omega_{c}h^2$,  effective acoustic scale $\theta_{\rm MCMC}$,  scalar power spectrum log amplitude $\mathrm{ln} (10^{10} A_s)$ and tilt $n_s$.   We fix the neutrino mass to one massive species with $m_{\nu}= 0.06$eV. 

Following Ref.~\cite{Mortonson:2008rx}, we adopt flat priors on the PC amplitudes  within the bounds of physicality for
the ionization history.  By requiring that $0 \le x_e \le x_e^{\rm max}$, each PC amplitude individually must lie in the range
$m_a^{-} \le m_a \le m_a^{+}$ where
\begin{equation}
m_a^{\pm} = \int_{z_{\rm min}}^{z_{\rm max} } dz \frac{S_a(z)[x_e^{\rm max} -2 x_e^{\rm fid}(z)]
\pm x_e^{\rm max} | S_a(z)|}{2(z_{\rm max}-z_{\rm min})},
\label{eq:individualprior}
\end{equation}
and by requiring
\beq
\int_{z_{\rm min}}^{z_{\rm max}} dz (x_e(z) - x_e^{\rm fid})^2 \leq (\zmax - z_{\rm min})(x_e^{\rm max} - x_e^{\rm fid})^2,
\eeq
 their joint variation must satisfy
\begin{equation}
\sum_{a} m_a^2 \le (x_e^{\rm max}-x_e^{\rm fid})^2.
\label{eq:jointprior}
\end{equation}
Here we take $x_e^{\rm max} = 1+f_{\rm He}$ for fully ionized hydrogen and singly ionized helium.

These are necessary but not sufficient conditions for physical ionization models $0\leq x_e(z)\leq1+f_{\rm He}$ for at most singly ionized helium. 
This is because  the omitted components impact physicality even though they do not affect the observables.  
 On the other hand, when testing
physical models for reionization with the constraints on the PC amplitudes, neither this prior nor
the unphysical models that it still allows matter \cite{Heinrich:2016ojb}. 

\begin{figure}
                    \includegraphics[width=\columnwidth]{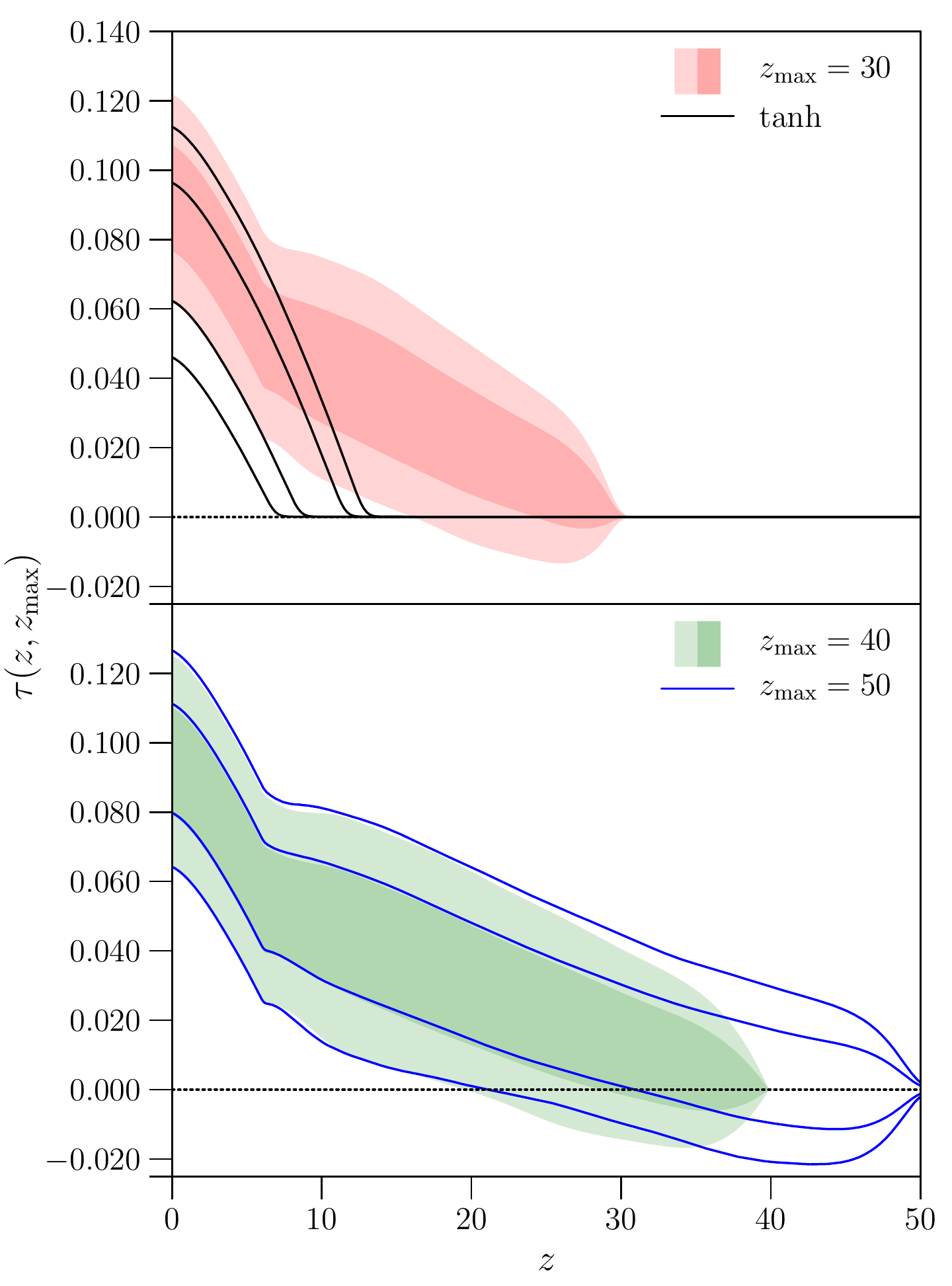}
          \caption
          {Cumulative optical depth $\tau(z,z_{\rm max})$ constraints (68\% and 95\% CL) for the various cases. Top: tanh (black lines) and PC chains with $\zmax = 30$ (red bands). The functional form of the tanh model strongly disfavors optical depth at $z\gtrsim 15$ due to constraints on the total $\tau$ whereas PC constraints for $\zmax=30$ allow and even favor these contributions at $95\%$ CL. Bottom: same for $\zmax= 40$ (green bands) and 50 (blue lines). Extending to higher redshift reveals a small amount of optical depth at $z = 30$ ($\sim 1 \sigma$ excess) missed in $\zmax = 30$ analysis. This moves the $2\sigma$ point of excess from $z\sim 15$ in $\zmax = 30$ to about $z \sim 20$ in both the $\zmax= 40$ and 50 chains. The results are stable between the last two because of negligible contribution from $z\gtrsim 40$.  }
              \label{fig:tau_gtz_tanh_vs_zmax30}
\end{figure}

\begin{table}[]
\centering
\caption{Total and high redshift optical depth constraints for different analyses of Planck 2015 data. }
\label{tab:tau}
\begin{tabular}{| c  |c |@{\hskip 0.06in} c@{\hskip 0.1in}  c @{\hskip 0.1in} c|}
\hline Model  &    $\zmax$         & $\tau(0,z_{\rm max})$             & $\tau(15, z_{\rm max})$  & $\tau(30, z_{\rm max})$ \\ \hline
tanh  &   ...   & 0.079 $\pm$ 0.017 & ... & ... \\ 
PC &    30    & 0.092 $\pm$ 0.015 & 0.033 $\pm$ 0.016     & ...                  \\  
PC &    40    & 0.095 $\pm$ 0.016 & 0.039 $\pm$ 0.017    & 0.013 $\pm$ 0.014                 \\  
PC &    50    & 0.096 $\pm$ 0.016 & 0.040 $\pm$ 0.018   & 0.016 $\pm$ 0.014                 \\ \hline 
\end{tabular}
\end{table}

Since the meaning of the $m_a$ amplitudes themselves change with $z_{\rm max}$, we instead
show the more robust and simple to interpret cumulative optical depth  $\tau(z,z_{\rm max})$ 
(see Eq.~\ref{eq:cumtau}).
In Fig.~\ref{fig:tau_gtz_tanh_vs_zmax30}, we plot the 68\% and 95\% confidence bands for the tanh model and PCs with $z_{\rm max} = 30,40,50$. 
For the tanh model, optical depth at $z> 15$ is strongly forbidden whereas in the
other cases it is preferred at more than the 95\% CL.   As discussed in Ref.~\cite{Heinrich:2016ojb}, in the tanh model constraints on the total optical depth forbid a high redshift 
component due to its functional form not due to other properties of the data: ionization at 
high redshift must be accompanied by full ionization at all lower redshift.   Relaxing this restriction with PCs out to  $\zmax=30$ uncovers a $2\sigma$ preference of
optical depth at $z> 15$ \cite{Heinrich:2016ojb}.  Here, we show that these results
are robust to further extending $\zmax$.   In fact, the significance for $z > 15$ in the $\zmax = 30$ chains increases slightly from $2.1\sigma$ to $2.3\sigma$ in the $z_{\rm max} = 40$ and 50 chains and there is a $\sim 1\sigma$  excess at $z\gtrsim 30$ (see Table~\ref{tab:tau}). The point of $2\sigma$ excess moves from $z \sim 15$ to $z \sim 20$ and is stable between $z_{\rm max} =  40$ and 50. This stabilization is related to the fact that there is no preference for ionization beyond $z \sim 40$. 

Despite differences in redshift range ($\zmax=30-50$), number of PCs (5-7) and fiducial models
around which the PCs are built, the constraints on the total optical depth remain consistent within the errors 
 (see Table~\ref{tab:tau}).   This is especially notable since as the number of PCs and $z_{\rm max}$ increases, there is a
larger prior volume for raising $\tau$ at high redshift compared with a choice of prior that
is flat in $\tau$. Yet, unlike  the WMAP data \cite{Mortonson:2008rx}, the Planck 2015 data themselves do not allow such variations (see Appendix \ref{sec:prior} for an extensive discussion).   Conversely we caution the reader that PC constraints on $\tau$ are weak enough to still be subject to implicit and
explicit priors from a given reionization model and should be interpreted in a model context for its impact  on ionization source or other cosmological parameters.   If these high redshift degrees of freedom are disfavored by the model,
e.g.~as in a population II scenario of reionization, then the PC constraints should be 
projected onto the model space with appropriate priors on the model parameters using
the procedure  in Ref.~\cite{Heinrich:2016ojb} as illustrated in \cite{Miranda:2016trf}. 

\begin{figure}
          \includegraphics[width=1\columnwidth]{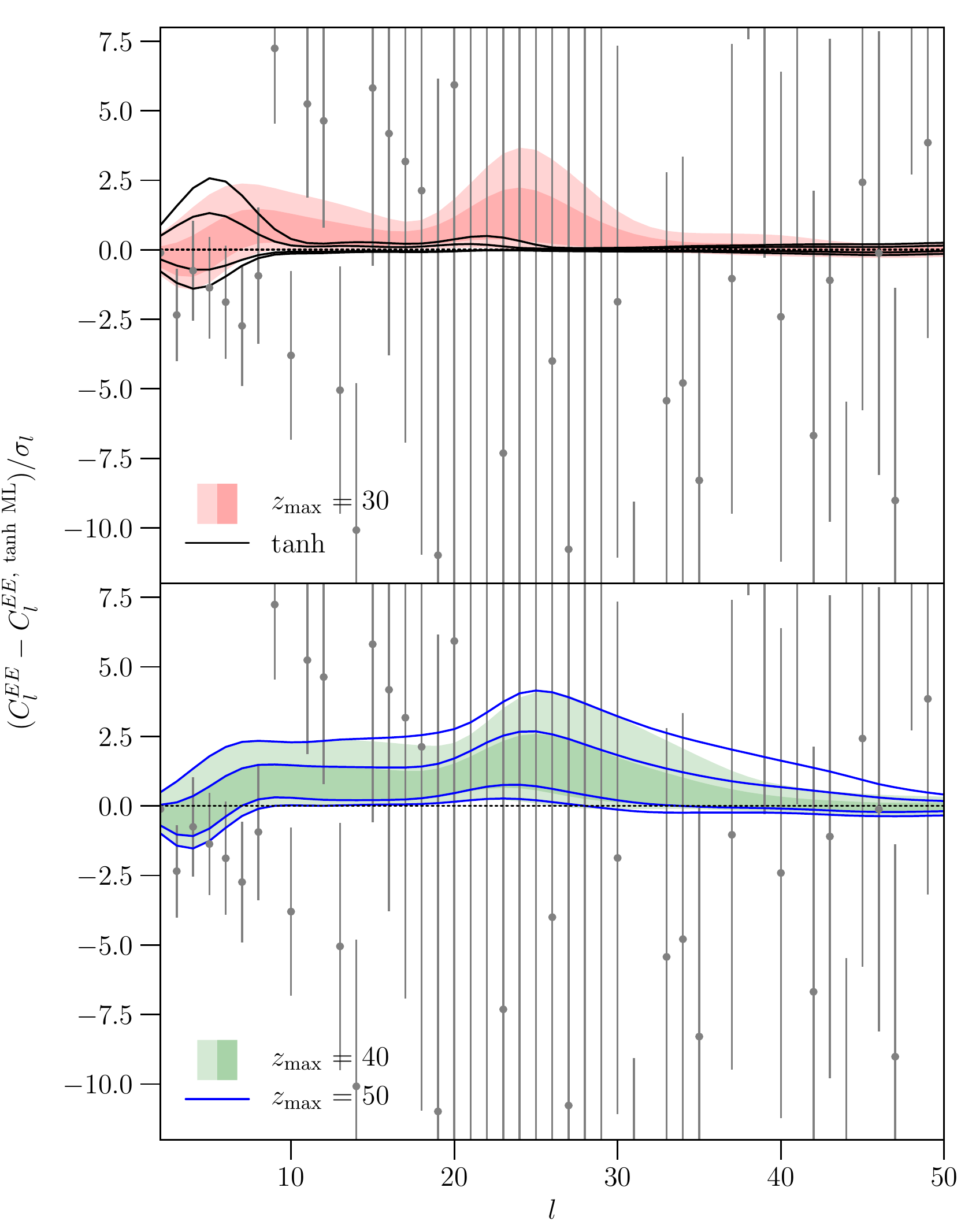}
          \caption
          {$EE$ data residuals and model posteriors with respect to the tanh maximum likelihood model scaled by the cosmic variance per multipole of the $\zmax=50$ fiducial model. Top: 68\% and 95\% CL posterior constraints from tanh (black lines) and PC chains with $\zmax = 30$ (red bands).  The high data point at $\ell=9$ and subsequent multipoles
          are generally better fit by PCs which allow for high redshift ionization. Bottom: same for
          $\zmax= 40$ (green bands) and 50 (blue lines).  Further increasing $\zmax$ 
          produces similar fits around $\ell \sim 10$ and allow more freedom at $\ell \sim 15-20$ 
          and $\ell \gtrsim 30$ where the data do not significantly constrain the models. }  \label{fig:ClEE_post}
\end{figure}

\begin{figure}
          \includegraphics[width=1\columnwidth]{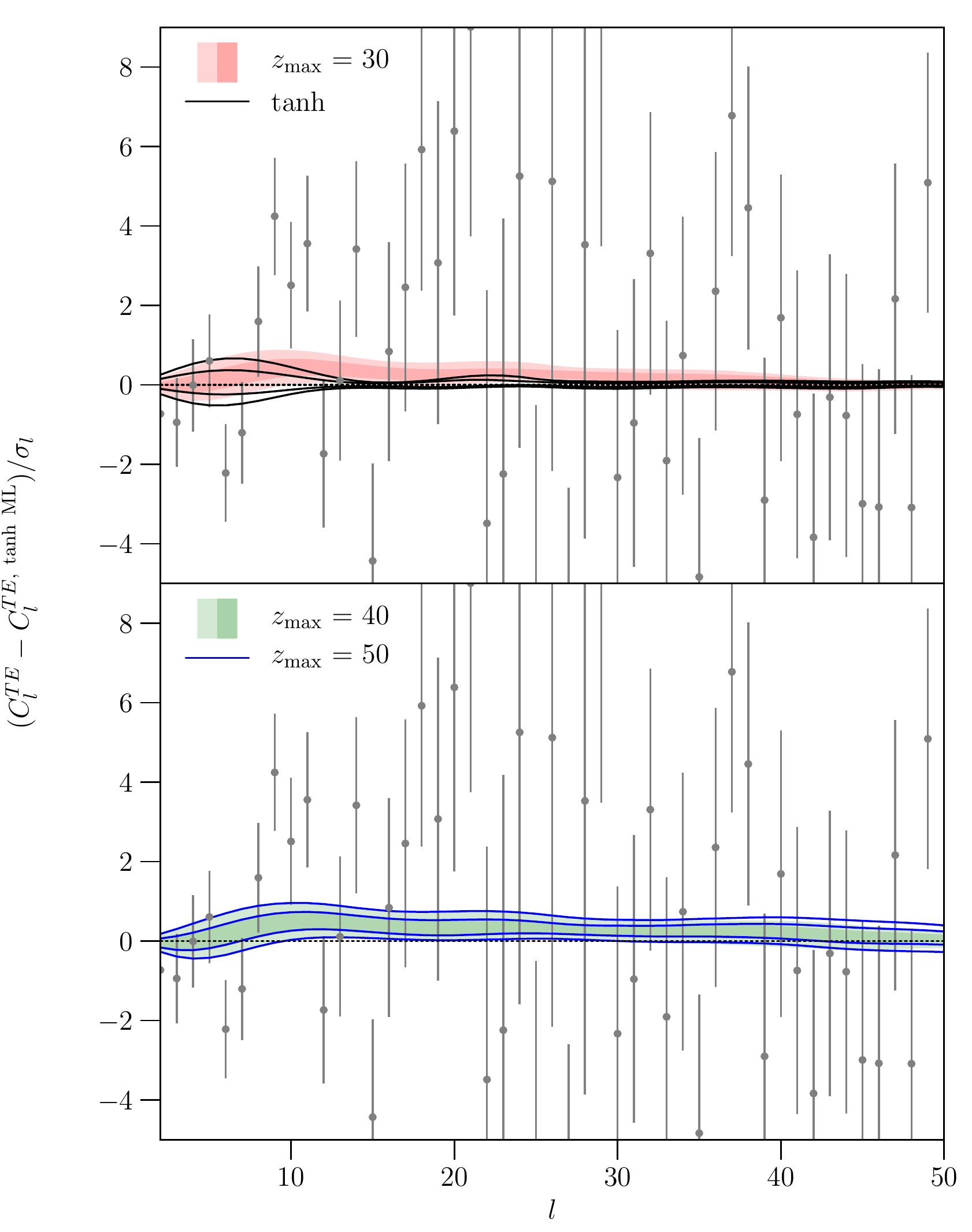}
          \caption
          {$TE$  data residuals and model posteriors as in \reffig{ClEE_post}.   As with $EE$, the high data points around $\ell \sim 10$ allow for and favor high redshift ionization that is absent in the tanh model.    The results in the well-constrained regions are again
          stable to $\zmax$.  
          }  \label{fig:ClTE_post}
\end{figure}

To better understand the origin of the robustness of the high-$z$ ionization constraints to $z_{\rm max}$,
we can compare the Planck $EE$ and $TE$ power spectrum data to the posterior probability distributions implied by the ionization constraints.
In Figs.~\ref{fig:ClEE_post} and~\ref{fig:ClTE_post} respectively we show the
68\% and 95\%  CL ranges  for the tanh model compared to $z_{\rm max}=30$ and
$z_{\rm max}=40$ to $z_{\rm max}=50$.  We see that the main aspect of the data that
tanh has difficulty explaining are the low $EE$ polarization points at $\ell \le 8$ compared to the
relatively high points that follow, especially $\ell=9$ \cite{Aghanim:2015xee}.   The tanh ionization history cannot raise the
latter without violating the constraints of the former.    Furthermore the significance of this point in $TE$ is also increased due to a slightly low fluctuation in $TT$ there.  On the other hand all 3 $z_{\rm max}$ cases 
produce very similar posteriors in the $\ell \lesssim 10$ range.  

As we have seen in Fig.~\ref{fig:dClEE_dxe}, raising the power at $\ell \sim 10$ requires ionization at $z>10$, but a wide range of redshifts have a similar impact there. The fit to the  Planck data  improves only very slightly with $\zmax=40$ by allowing more power between $\ell \sim 10-15$.  
 Instead, the difference between the various $\zmax$ cases appears mainly at $\ell \gtrsim 15$ where the data do not yet significantly constrain the models.    
 In particular
the power spectra posteriors of $\zmax = 40$ become more significantly distinguishable from $\zmax=30$ with a larger vertical range mainly at $\ell \sim 15-20$ (see Fig.~\ref{fig:ClEE_post}).
For the $\ell \gtrsim 30$ region, while the response functions in Fig.~\ref{fig:dClEE_dxe} imply systematically larger power there as a function of redshift, and the posteriors of Fig.~\ref{fig:ClEE_post} are different between all three $\zmax$, the Planck errors there are too large to have an impact on the models.  Since these posterior differences are
above the cosmic variance limit, they are in principle testable with future high precision
ground or space based polarization measurements, especially should the true model have little ionization above $\zmax=30$. 

Correspondingly the maximum likelihood models show improvements that support a $\sim 2\sigma$ preference for high redshift ionization mainly in
the $\ell < 30$ part of the likelihood.   Note that the likelihood $\mathcal{L}$ reflects the goodness of fit of a particular set of parameters and is not sensitive to the prior chosen for the parameters unlike
posterior constraints on $\tau$ and power spectra.  
Hence the preference for high-$z$ ionization is not simply due to the difference between flat priors on PCs and flat priors in $\tau$ 
as we explicitly demonstrate in Appendix~\ref{sec:prior}.  In Table~\ref{tab:best-fit} we list the improvements of the best fit PC models over
the tanh cases: for $\zmax=\{ 30,40,50\}$, $2\Delta \mathrm{ln}\mathcal{L} =  \{5.5,5.9,6.1\}$ respectively (see also \cite{Hazra:2017gtx}).
Note that for $\zmax=50$, this final further improvement is negligible despite having two additional PC parameters.

Robustness to $\zmax$ of these improvements further indicates that this mild preference for
reionization histories beyond the tanh model really reflects a single aspect of reionization models: the presence of $z>10$ ionization.   Hence this improvement should be considered
as due to a single parameter in spite of the larger number of PCs and the wide range in
redshift allowed in the analysis.   We provide a concrete illustration of this single parameter in Appendix \ref{sec:highz_model}.
  The number of PCs is  chosen for completeness  to the
cosmic variance limit  not
because the additional parameters are required by the Planck data.   This has the benefit that 
constraints on any model of reionization out to the same $\zmax$ can be directly obtained
from the PC posteriors for Planck and for any future CMB experiment.  

Conversely the $2\Delta \mathrm{ln}\mathcal{L} =  5-6$ improvement does not provide a highly statistically
significant detection of high-$z$ ionization on its own.   In the context of models like the tanh case which
disfavor high-$z$ ionization by an explicit or implicit modeling  choice (see also \cite{Villanueva-Domingo:2017ahx}), this difference would simply
be attributed to a statistical or systematic fluctuation.   For models that do allow it (e.g.~\cite{2012ApJ...756L..16A}),
the high-$z$
ionization  window is open and even mildly favored in the Planck 2015 polarization data
due to features in the data at $\ell \sim 10$.

\begin{table}[]
\centering
\caption{ Improvement of the Planck 2015 maximum likelihood  $2\Delta \mathrm{ln} \mathcal{L}$ 
 in different classes with respect to the tanh best-fit. } \label{tab:best-fit}
\begin{tabular}{| c  |c |c |}
\hline Model  &    $\zmax$ & $2\Delta\mathrm{ln}\mathcal{L}$ \\ \hline 
tanh  &   ...   & 0.0 \\
PC &    30    & 5.5 \\
PC &    40    & 5.9 \\
PC &    50    & 6.1 \\ 
two step & ... & 5.3 \\  \hline
\end{tabular}
\end{table}

\section{Conclusion}\label{sec:discussion}

In this work, we explored the relationship between  signatures of high redshift ionization in CMB power spectra and features in the Planck 2015 LFI polarization data.    We extend previous work
by analyzing a complete parameterization of reionization to the cosmic variance limit 
out to $\zmax=40$ and $\zmax =50$ and identifying the specific aspect of the data that
drive the constraints.  

 In a previous work with $\zmax = 30$, we found a $2.1\sigma$ excess in cumulative optical depth at $z \sim 15$. In the $\zmax = 40$ and 50 analyses, this excess is stable and in fact, slightly enhanced to $2.3\sigma$ because of the additional freedom to have optical depth
 at $z>30$. Beyond $z\approx40$, however, there is no preference for finite ionization.
 
The origin and robustness of these results is related to data in the $\ell \sim 10$
region of the $EE$ and $TE$ power spectra.  These data points are generally higher than can be accommodated by a reionization history with only $z<10$ ionization as in the steplike
tanh model.   Specifically, such models have difficulty simultaneously fitting the low
power  at $\ell \leq8$ and abruptly higher power at $\ell=9$ simultaneously.
Allowing for partial ionization out to $z>10$ through the PC basis can better 
accommodate the data.   On the other hand the $\ell \lesssim 10$ region, which carries most
of the signal to noise for Planck, cannot be used to discriminate the high redshift range
further due to a near degeneracy in observational effects.   Instead, differences 
appear at $\ell \sim 15-20$ and also
at $\ell \gtrsim 30$.  Though significant at the cosmic variance limit, for the Planck data the whole $\ell>10$ regime only provides a mild  preference for  $z\lesssim40$. 

These results are also consistently reflected in the moderate improvement of the maximum likelihood once
high redshift ionization is accommodated with PC parameters: $2\Delta \mathrm{ln}\mathcal{L}$ of 5.5, 5.9 and 6.1 for $\zmax = 30$, 40 and 50 respectively.  
These improvements are independent of the prior chosen for the parameters
which is important since the PC approach  entails a larger parameter volume for models
with high redshift ionization than a flat prior on the total optical depth.   This difference in prior volume is not relevant because the Planck 2015 data
are more informative than the PC priors as we explicitly show in the Appendix.

The robustness
of the results to $\zmax$ also indicates that these improvements should be viewed as 
originating from a single aspect of reionization models: the ability to accommodate
$z>10$ ionization.  The much larger number of PC parameters used in the actual analysis
is chosen for completeness to the cosmic variance limit and not because they are required by the Planck data.  These  additional parameters are also not just artificially increasing the likelihood by 
fitting discrete noise fluctuations as the lack of further improvement of the likelihood with
increased PC number or $\zmax$ shows.  In the Appendix we provide a concrete example where the improvements are encapsulated
into a single parameter for the high $z$ optical depth.

On the other hand, even if due to one single effective parameter, improvement at this level does
not constitute strong evidence for high redshift ionization on its own.   The benefit of the
PC approach being complete to the cosmic variance limit is that any reionization model can be projected onto this basis without loss of information and interpreted with physically motivated priors.    
For models which accommodate high redshift ionization, a $z>10$ component is clearly allowed and even mildly preferred.  Constraints on the total optical depth assuming a tanh model should not be used to exclude high redshift ionization in these models.
Within a model class that forbids
high redshift ionization, the poor fits would be attributed to a $2\sigma$ statistical fluctuation or
systematic effects.  Since the data are far from the cosmic variance limit, better measurements
especially at $\ell =15-20$ and at $\ell >30$ can potentially decide this issue due to the
excess power that high redshift ionization predicts there. 

Indeed, the Planck team has done substantial work following the 2015 release on
removing systematics from the low multipole measurements from the much more sensitive 
HFI measurements \cite{Aghanim:2016yuo}. In particular, the 2016 intermediate release included a tanh analysis reporting a lower value of $\tau = 0.055 \pm 0.009$. As the tanh analysis provides essentially a low-$z$ estimate of optical depth, the high-$z$ component still remains unconstrained in this analysis. 
After this work was largely complete, Ref.~\cite{Millea:2018bko} analyzed the 2016 intermediate release with principal components and
found tighter constraints on the high-$z$ component.  However, these data are still proprietary and the final full data set will soon be released publicly.
In this context, our study of how features in the data translate to features or
constraints on reionization at high redshift is valuable in determining the optimal strategy for
extracting the most information from this and any future CMB data set. 

\appendix

\section{Optical depth parameters}
\label{sec:appendix}

\subsection{How to (and not to) flatten the $\tau$ prior}
\label{sec:prior}

After this work was largely complete, a concern was raised in Ref.~\cite{Millea:2018bko} (hereafter MB18) regarding the $\tau$ prior implied by projecting the multidimensional flat PC prior onto that single dimension. MB18 claimed that since this prior is not flat in $\tau$, it leads to a bias in the PC result.
 In addition, MB18 proposed flattening the prior by multiplying by its inverse point-by-point in the multidimensional PC space
 and claim to show that flattening removes 
a large fraction of the shift of the $\tau$ posterior between tanh and PC results.  We point out in this appendix that this logic is flawed and that even
with a flat prior in $\tau$ the preference for a high redshift component remains.

When  applied to  datasets that are actually informative in the original multidimensional space, the MB18 approach to flattening actually introduces rather than removes bias.  The problem is that
it weights regions of the parameter space that are allowed by the original prior but
rejected by the data and thus creates a new prior that is {\it not} locally flat across the region of support
for the posterior.  Their technique should only be applied to data that is  informative solely on  $\tau$, or more generally the parameter whose projected prior is flattened,  and not on any of the other  parameters in the space.

\begin{figure}
          \includegraphics[width=1\columnwidth]{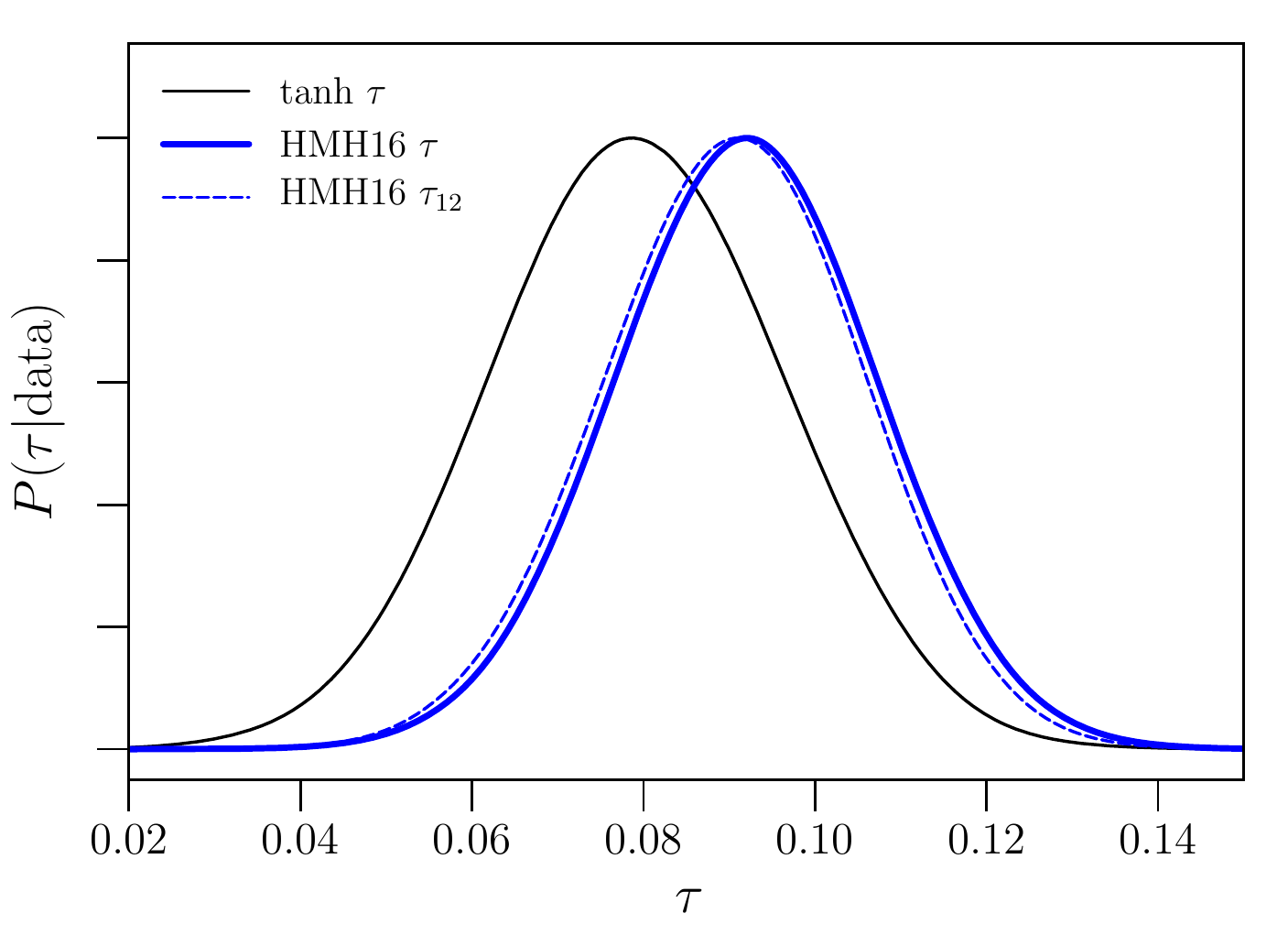}
          \caption
          {Posteriors of $\tau$ constructed from the complete 5 PCs (solid blue) vs.~the first two PCs $\tau_{12}$ (Eq.~\ref{eq:tau12}) (dashed blue), both constructed with the physical prior of HMH16. The difference between them are very small compared to the relevant shifts discussed in this appendix, 
          especially the shift between the tanh (black) and PC results. }  \label{fig:tau12_vs_tau}
\end{figure}

 To illustrate this explicitly in the context critiqued by MB18, we consider the case of the $\zmax=30$ analysis from HMH16 \cite{Heinrich:2016ojb}.  The total optical depth is a linear combination of contributions from the individual PCs
 \begin{equation}
 \tau =\tau_{\rm fid} +  \sum_a m_a \tau_a
 \end{equation}
 where $\tau_a$ is the optical depth for a unit amplitude PC and $\tau_{\rm fid}$ is the optical depth of the fiducial model.    This linearity already indicates that if a given model parameterizes a linear trajectory in the
 PC space, then a flat prior on $m
 _a$ produces a flat prior in $\tau$ (see HMH16 for the tanh trajectory).  This is the key to understanding why the flat prior on $m_a$ is compatible
 with a prior on $\tau$ that is {\it locally} flat.  
 
 \begin{figure}
          \includegraphics[width=0.9\columnwidth]{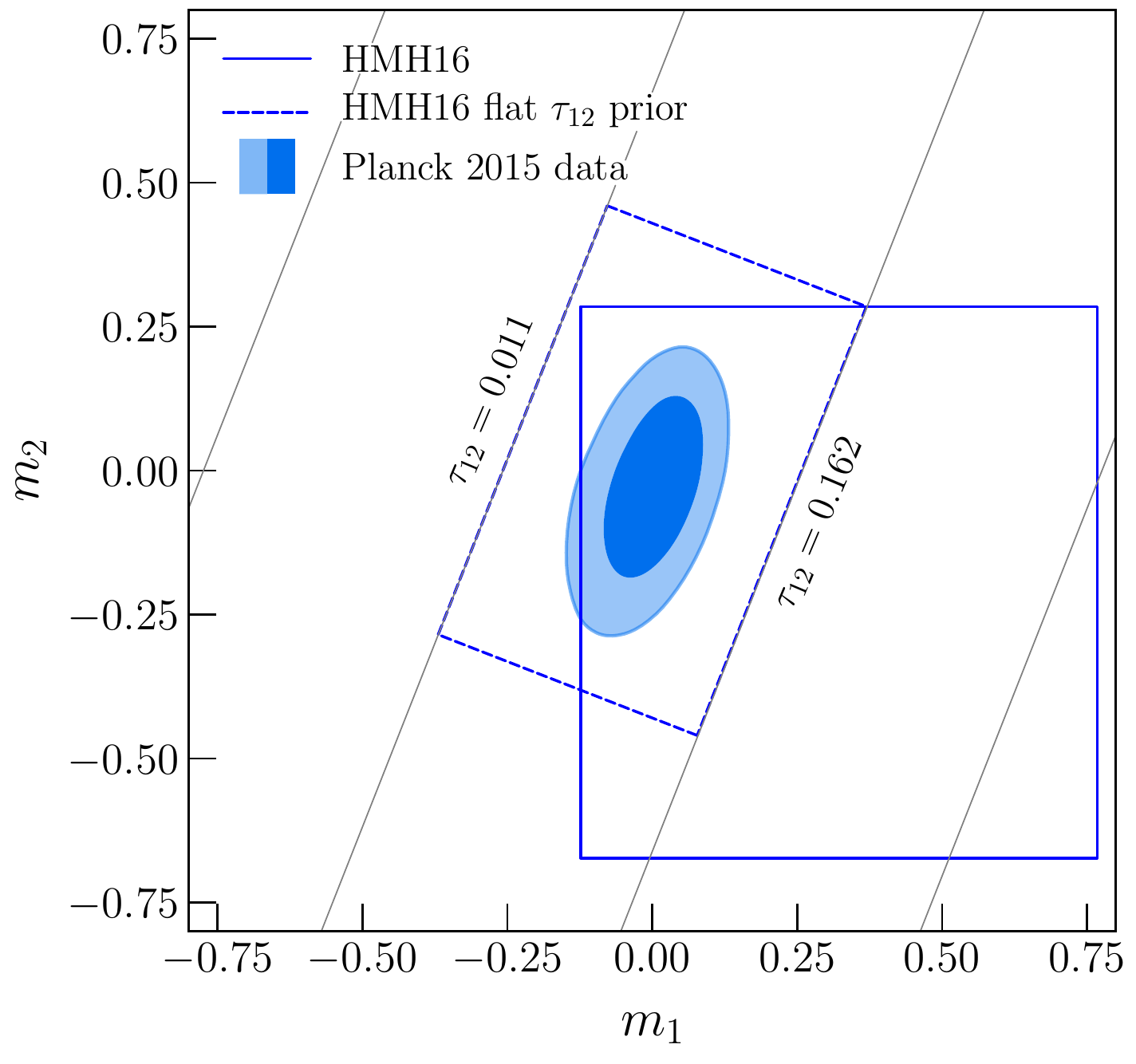}
          \caption
          {Priors on the $m_1-m_2$ parameter space: the original HMH16 prior (inset square, solid lines) and a prior that is flat in
          both $\tau_{12}$ and $m_1-m_2$ (rotated rectangle, dashed lines) whose sides are aligned with lines of
          constant $\tau_{12}$ (light gray lines).   The HMH16 prior allows more parameter space at high vs low $\tau_{12}$ but this
          is not relevant since the data constraints (ellipses) exclude this region.  In the allowed region, the only difference is that the HMH16 prior clips the low $m_1$ edge of the allowed region due to physicality and the assumption that reionization occurs at
          $z\ge 6$. }  \label{fig:prior_box}
\end{figure}

 In the HMH16 case, almost all of the PC contribution to $\tau$ comes from the first two components
 \beq
\tau_{12} =\tau_{\rm fid}+ m_1 \tau_{1} + m_2 \tau_{2} \approx \tau.
\label{eq:tau12}
\eeq
 In Fig.~\ref{fig:tau12_vs_tau} we show that the posteriors for $\tau_{12}$ and $\tau$, both with the original HMH16 priors
 from Eq.~(\ref{eq:individualprior}) and (\ref{eq:jointprior}), differ very little compared to the shift from the tanh model and likewise to any other relevant shifts discussed below.   
 
 \begin{figure}
          \includegraphics[width=1\columnwidth]{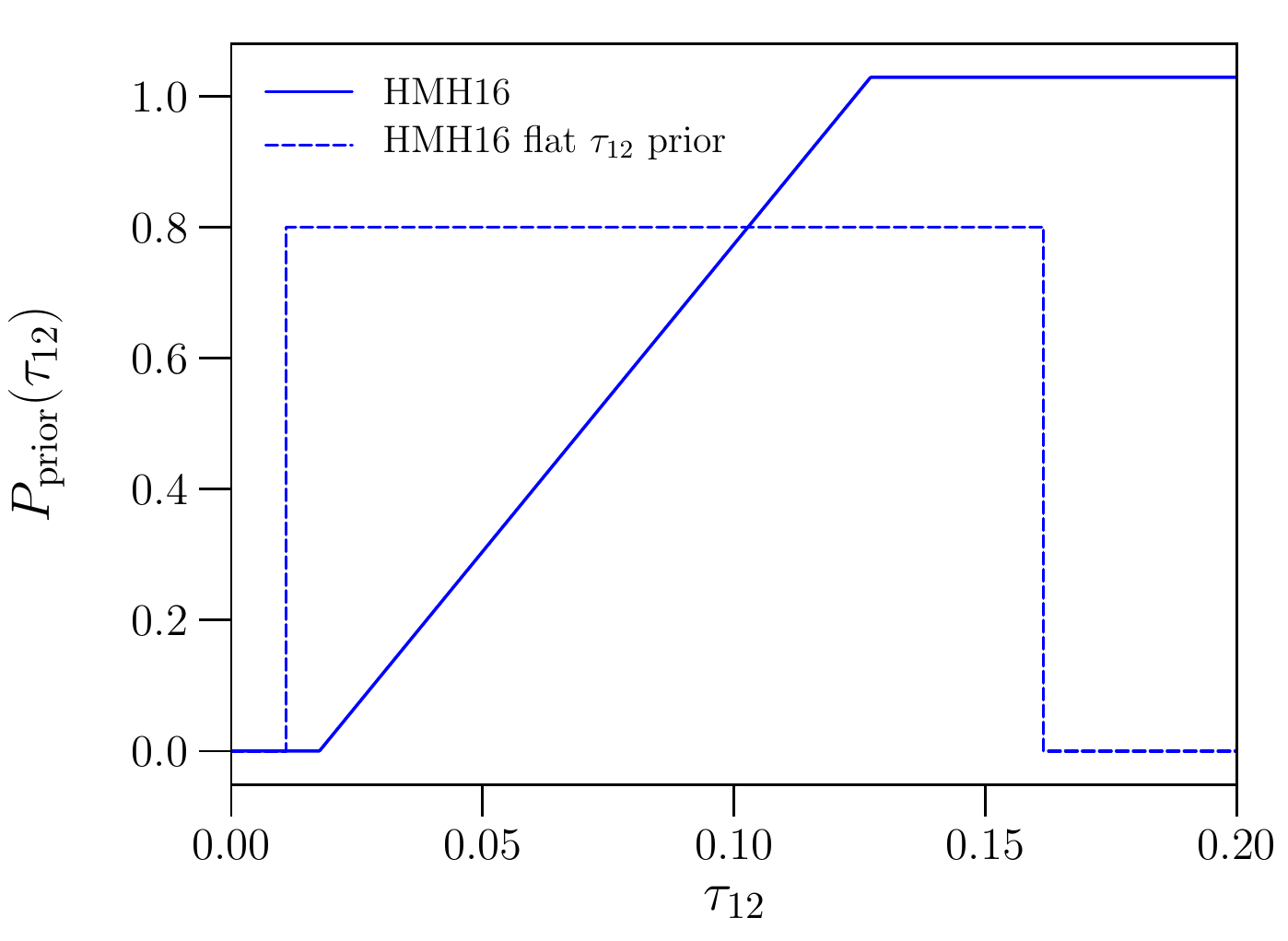}
          \caption
          {The integrated $\tau_{12}$ priors implied by the $m_1-m_2$ priors shown in Fig.~\ref{fig:prior_box}: 
          the HMH16 prior (solid line) is not flat due to the larger prior volume at high $\tau_{12}$ whereas the flat $\tau_{12}$ prior (dashed-line)
          is a flat top hat by construction.    These two priors appear in projection to be quite different but locally in the region allowed by the 
          data in Fig.~\ref{fig:prior_box} are almost the same.  
         } \label{fig:prior_plot}
\end{figure}

 By focusing the analysis on $\tau_{12}$, we can easily analyze purported biases from the PC priors.  In Fig~\ref{fig:prior_box}, 
 we show the HMH16 priors on the $m_1-m_2$ space as the inset square (solid lines).   
Lines of constant $\tau_{12}$ are diagonal lines (light gray lines). 
 As in MB18, we can project the HMH16 prior onto the $\tau_{12}$ direction.  The result is the  non-flat prior in $\tau_{12}$ 
 shown in Fig.~\ref{fig:prior_plot} (solid line).   This shape can be easily understood as the length of the line segments of the
 constant $\tau_{12}$ lines that lie within the HMH16 prior box in Fig.~\ref{fig:prior_box}.   
 Note that the prior in  Fig.~\ref{fig:prior_plot} linearly rises with  $\tau_{12}$ in the region of support of the posterior in 
 Fig.~\ref{fig:tau12_vs_tau} leading MB18 to claim a biased prior.
However  the Planck 2015 data is more informative than the prior in both $m_1$ and $m_2$ (Fig.~\ref{fig:prior_box}, ellipses) making 
 this elongation of the line segments at high $\tau_{12}$ irrelevant for the posterior.  
 
\begin{figure}
          \includegraphics[width=1\columnwidth]{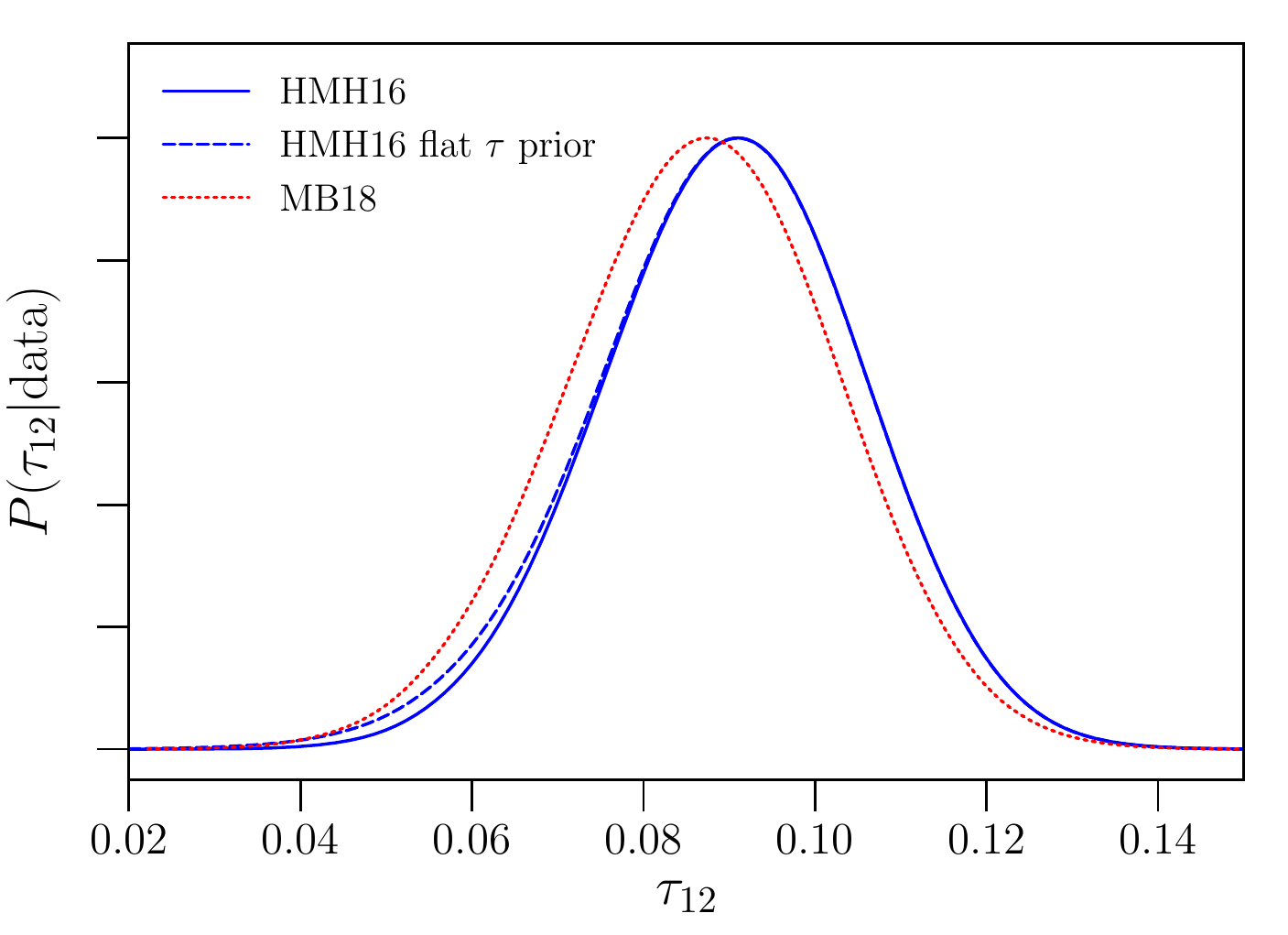}
          \caption
          {The $\tau_{12}$ posterior obtained using the HMH16 prior (solid blue) is nearly identical to that of the flat $\tau_{12}$ prior (dashed blue) 
          except for the low-$\tau_{12}$ tail.    This difference is the result of assuming reionization occurs at $z\ge 6$ and the physicality constraints of HMH16.  
           The flattening procedure of MB18, on the other hand, introduces a bias shifting the entire posterior toward lower-$\tau_{12}$ due to the weight
           it places on regions allowed by the prior but forbidden by the data. }  \label{fig:flattening_vs_flat}
\end{figure}

 By following the MB18 procedure of inverting this prior by multiplying each point in the PC sample by $P^{-1}_{\rm prior}[\tau_{12}(m_1,m_2)]$,
 we reproduce their results of a shift to lower $\tau_{12}$ for the posterior in Fig.~\ref{fig:flattening_vs_flat}.  However this result occurs because the inversion weights the prior volume in regions
 that are not supported by the data and hence produces a bias against high $\tau_{12}$ in the region that is supported by the data, i.e.\ it makes the
 $\tau_{12}$ prior globally flat when integrated over the whole prior volume by making it locally {\it not flat} across the allowed region.
 
 To make this point even more clear,  we can explicitly construct a prior that is locally flat in both $m_1-m_2$ and $\tau_{12}$.  Consider
 the rotated box shown in  Fig~\ref{fig:prior_box} (dashed lines).    Now the edges of the box are lines of minimum and maximum $\tau_{12}$ and the 
 parameter orthogonal to $\tau_{12}$.  As shown in Fig.~\ref{fig:flattening_vs_flat} (dashed line), this new prior is explicitly flat in $\tau_{12}$ between
 its minimum and maximum.   In Fig.~\ref{fig:flattening_vs_flat}, we show that the posterior from around its maximum out to high $\tau_{12}$ shifts negligibly between
 this new flat $\tau_{12}$ prior and the original HMH16 prior.   The main impact is a slight enhancement at very low $\tau_{12}$ over the HMH16 
 results.   This is because the HMH16 prior takes into account
 physicality constraints and restricts hydrogen to be fully ionized at $z\le 6$.   As shown in Fig.~\ref{fig:prior_box}, this prior slightly clips 
 the 95\% CL region of the PC posterior at low $m_1$.   This effect is qualitatively similar to what would occur in a tanh model if we restrict the redshift of
 reionization to $z >6$  (see also Fig.~\ref{fig:tanh_highz}).  Note also that this effect is much smaller than the bias that results from applying the MB18 procedure.  We have also explicitly
 checked that changing the size of the rotated prior box does not affect the posterior within its region of support unless the box begins to 
 exclude this region.  
Similarly, the $\tau(15,\zmax)$ significance does not change much between the HMH16 and the flat $\tau_{12}$ prior result: $\tau(15,\zmax) = 0.033 \pm 0.016$ and $\tau(15,\zmax) = 0.032 \pm 0.018$ respectively.    
 
 We conclude that the MB18 procedure should not be applied to the Planck 2015 data set or any future measurement where the data is
 actually more informative than the physicality prior.  It is nonetheless equally important to bear in mind that the weak preference for high-$z$
 ionization in Planck 2015 data can still be outweighed by stronger explicit or implicit priors on the detailed astrophysics behind reionization as discussed in the main text.

\subsection{Adding a single high-$z$ $\tau$ parameter} \label{sec:highz_model}

Here we provide a concrete example of how the improved fit by the PC over tanh analysis, represents only one single aspect of the model -- high-$z$ optical depth. Even though in the PC analysis, each ionization history is parametrized by 5 mode amplitudes for $\zmax = 30$, the improvement
can be encapsulated into a single additional parameter.  This example demonstrates that the extra PC parameters in the analysis are included for completeness and are not just improving the likelihood incrementally by fitting random statistical fluctuations.

 \begin{figure}
          \includegraphics[width=1\columnwidth]{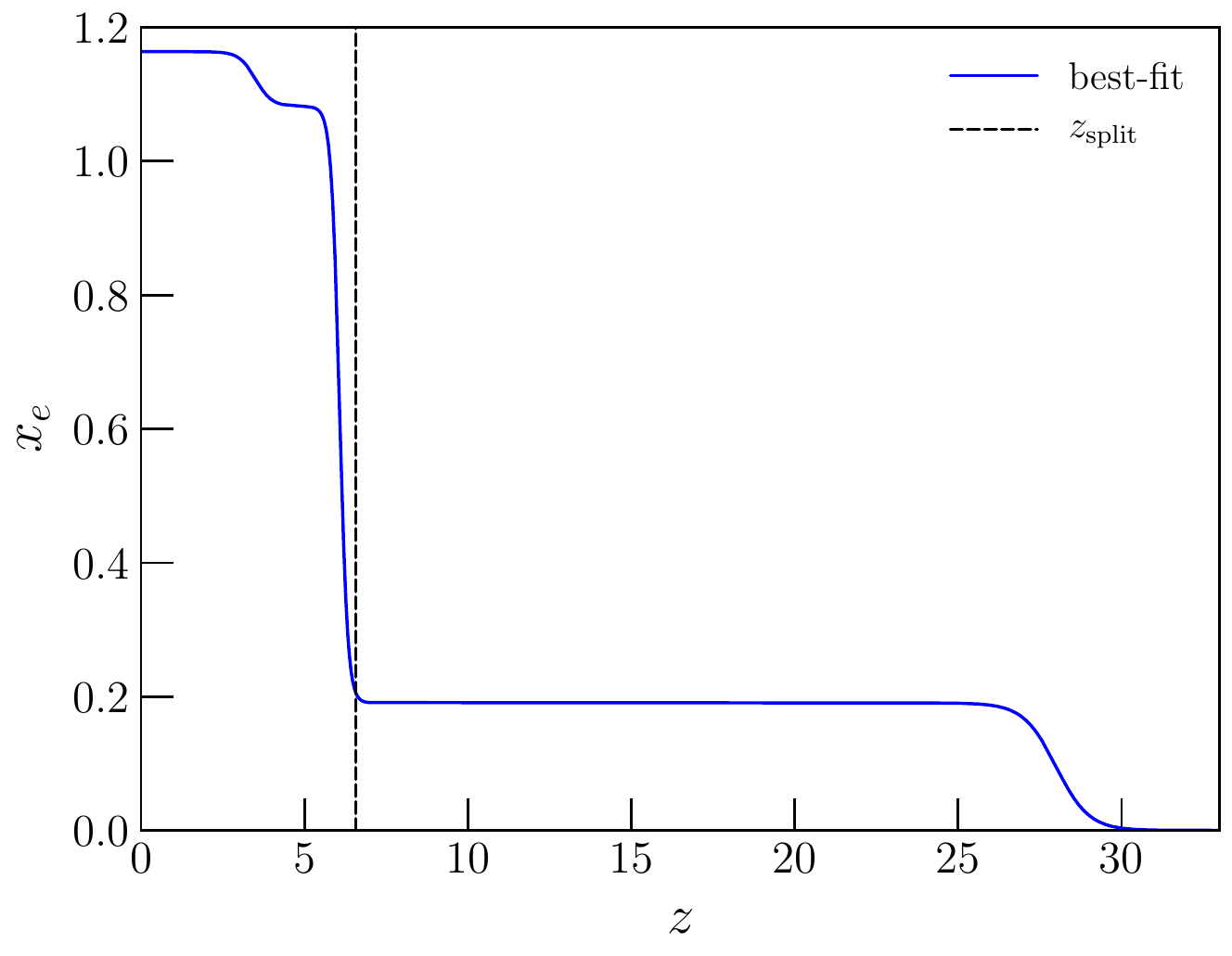}
          \caption
          {Best-fit two-step model (blue solid) with $\tau_{\rm lo} = 0.040$ and $\tau_{\rm hi} = 0.053$. The dashed black line shows the split used to define high vs low-$z$. }  \label{fig:two_step_model}
\end{figure}

Our example is a two-step model 
 \bea
x_e^{\rm true}&(z)&\,= \frac{1+f_{\rm He} - \xemin}{2}\left\{  1+ \tanh\left[ \frac{y(z_{\rm re})-y(z)}{\Delta y} \right] \right\} \notag \\
&+& \frac{\xemin - x_e^{\rm rec}}{2}\left\{  1+ \tanh\left[ \frac{z_{\rm t}-z}{\Delta z_{2}} \right] \right\} + x_e^{\rm rec},
 \label{eqn:tanh_highz}
 \eea
where $y(z)=(1+z)^{3/2}$, $\Delta y=(3/2)(1+z)^{1/2}\Delta z_1$, with $\Delta z_1 = 0.25$ instead of the usual $\Delta z_1 = 0.5$,
to provide sharper distinctions between the two steps.
   We choose the second step to have $z_{\rm t}=28$  and  $\Delta z_2 = 1.0$  to illustrate below
how the 5 PC analysis with $\zmax=30$ represents the same model.  Here $x_e^{\rm rec}$ is the ionization history from recombination only.
The canonical tanh model is essentially recovered in the limit $\xemin$ approaches the negligible $x_e^{\rm rec}$.  Therefore the double
step model adds a single parameter to control the high-$z$ ionization plateau for $z_{\rm re} \lesssim z \lesssim z_t$.   We show an example
of the two-step model in Fig.~\ref{fig:two_step_model}.

\begin{figure}[t]
          \includegraphics[width=0.95\columnwidth]{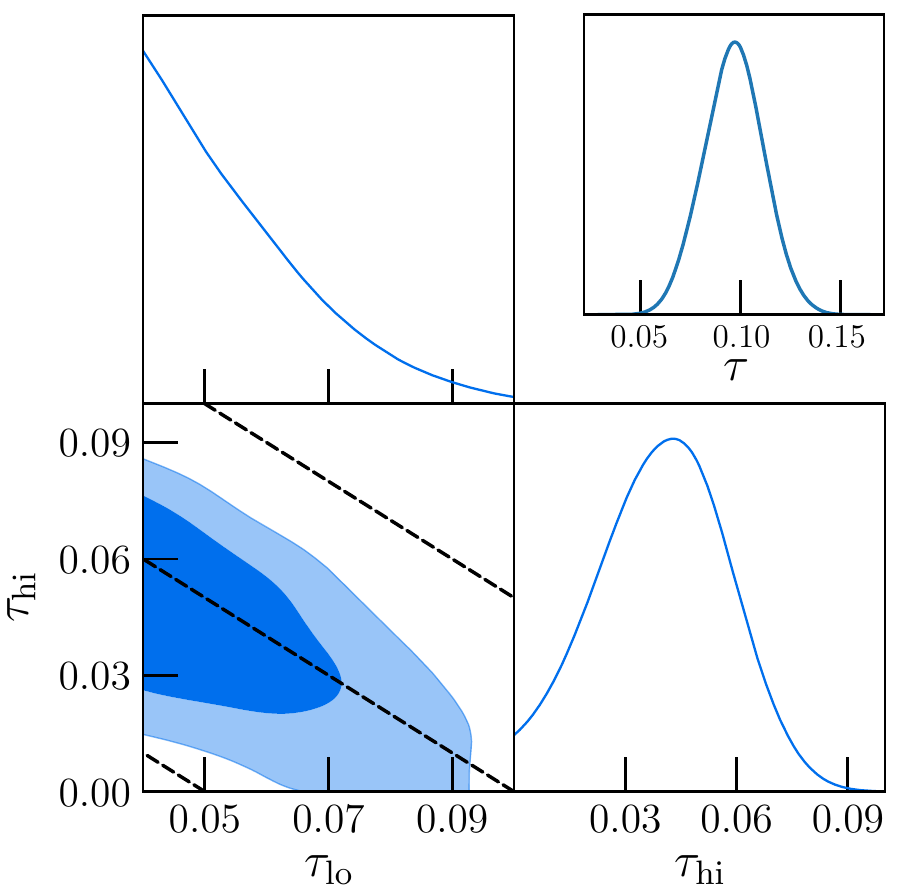}
          \caption
          {Low and high $z$ optical depth constraints from a two-step model that allows for high-redshift optical depth $\tau_{\rm hi}$. This component is detected at about $2\sigma$ significance. Dashed lines are lines of constant total $\tau = 0.05, 0.1$ and 0.15 and the implied $\tau$ posterior is shown in the upper right panel.
          Notice that the prior $\tau_{\rm low}\ge 0.04$ $(z_{\rm rei} \gtrsim 6$) suppresses the low $\tau$ tail by truncating its contributions (cf.~Fig.~\ref{fig:tau12_vs_tau}, tanh model).}
           \label{fig:tanh_highz}
\end{figure}

As in the previous section, we also illustrate our results using priors that are locally flat in $\tau$ but now divided into a low and high $z$ component by
the double step.  
We therefore take as MCMC parameters  $\tau_{\rm lo}(z_{\rm re},\xemin)$ and $\tau_{\rm hi}(z_{\rm re},\xemin)$, defined as Eq.~(\ref{eq:cumtau}) but with boundaries [0, $z_{\rm split}$] and [$z_{\rm split}$, $\infty$] respectively, such that the total $\tau = \tau_{\rm lo} + \tau_{\rm hi}$. We choose $z_{\rm split} = z_{\rm re} + 2\Delta z_1$ to be conservative on the preference for $\tau_{\rm hi}$ in the data (see Fig.~\ref{fig:two_step_model}).
We sample  the posterior of  $\tau_{\rm lo}$ and $\tau_{\rm hi}$ assuming a uniform prior in the intervals [0.04, $\tau_{\rm max}$] and [0, $\tau_{\rm max}$] respectively  where $\tau_{\rm max} = 0.35$; we also limit the total $\tau$ to $[0,\tau_{\rm max}]$.

Fig.~\ref{fig:tanh_highz} shows the posterior distribution of $\tau_{\rm lo}$ and $\tau_{\rm hi}$.  Here $\tau_{\rm hi}=0.040\pm 0.017$ and is  detected at about $2\sigma$ level similar to the PC result for $\tau(15,\zmax)$.   Likewise the best-fit two-step model ($\tau_{\rm lo} = 0.040, \tau_{\rm hi} = 0.053$) provides an improvement of
 $-2\Delta \ln {\cal L}  = 5.3$ over the best fit single-step tanh model whereas the best-fit PC parameters yields $5.5$ for $\zmax = 30$.
 In the two-step model this improvement comes from a single extra parameter $\tau_{\rm hi}$.  The posterior distribution of $\tau=\tau_{\rm lo}+\tau_{\rm hi}$ is
 also consistent with the shift between the tanh model and the PC posterior shown in Fig.~\ref{fig:tau12_vs_tau}.

One can further check that the $\zmax=30$ PC projection of the best-fit two-step model gives the same results as the direct method.
The PC parameters of the
model are  $m_{1..5} = \{0.010,-0.035 -0.009, -0.030, -0.005\}$.   The difference in the total optical depth between the direct result and the PC projection 
is $\delta \tau = 8\times 10^{-4}$, which is well below the errors in $\tau$.   Once this insignificant difference is accounted for by scaling $A_s$ to a fixed
$A_s e^{-2\tau}$, the difference in likelihood between the PC projected model and the direct model is $2\delta \ln {\cal L}=0.07$.  

This two-step example highlights the fact that we include the 5 PC parameters to  ensure that they can describe the polarization spectrum for
 any ionization history with $z<\zmax$ even if in the given model
there really is only one extra effective parameter causing the likelihood differences from the tanh model.  This supports our claim in the main text that
these likelihood differences do not simply represent the overfitting of random statistical fluctuations multipole to multipole with multiple parameters.
This example also addresses the question raised by 
MB18 of whether the remaining unphysical models allowed by the HMH16 priors bias results. 
 In this case, the prior is both locally flat in $\tau$ parameters and explicitly allows only models with positive ionization.

 \begin{acknowledgements}
 We thank Cora Dvorkin, Stefano Gariazzo, Olga Mena, Marius Millea, Vinicius Miranda and Georges Obied for useful discussions.  
WH was supported by  U.S. Dept. of Energy contract DE-FG02-13ER41958, NASA ATP NNX15AK22G and the Simons Foundation.  Computing resources were provided by the University of Chicago Research Computing Center through the Kavli Institute for Cosmological Physics at the University of Chicago. Part of the research described in this paper was carried out at the Jet Propulsion Laboratory, California Institute of Technology, under a contract with the National Aeronautics and Space Administration. 

\end{acknowledgements}

\bibliography{ReiP15Ext.bib}

\end{document}